**Evaluation of Atmospheric Electric Field as Increasing Seismic Activity Indicator on the example of Caucasus Region**

M.K. Kachakhidze[1], Z.A. Kereselidze[2], N.K. Kachakhidze[1]

[1] St. Andrew The First-Called Georgian University of The Patriarchy of Georgia, Tbilisi, Georgia

[2] Iv. Javakhishvili Tbilisi state university, M. Nodia Institute of Geophysics, Tbilisi, Georgia

*Correspondence to:* M. K. Kachakhidze    manana_k@hotmail.com

**Abstract**. The present paper deals with reliability of a gradient of atmospheric electric field potential as an indicator of seismic activity increase. With this in view, records of atmospheric electric field potential gradients of Caucasus region for 1953-1992 with respect to periods before average and large earthquakes, which took place in the same time interval, were considered.

It is worth to pay attention to the fact that the avalanche-like unstable model of fault formation based on theoretical model of self-generated seismo-electromagnetic oscillations of LAI system explains convincingly spectral succession of electromagnetic emission frequency of the periods preceding earthquakes.

**Introduction**

There are rather sufficient data about anomalous changes in atmospheric electric field gradients in the process of preparation of earthquakes in different regions. For example, similar research for Caucasus was implemented by us on the basis of multiyear data [Kachakhidze et al., 2009]. Although it is still considered that the available material is not enough to associate convincingly anomalous changes in atmospheric electric field with the depth processes going on in the area of incoming earthquake preparation.

On the other hand, in recent years it has become apparent that the earth VLF electromagnetic emission might be a result of increasing of seismic activity. It is logical to assume that this phenomenon as an independent electromagnetic indicator of the process of earthquake preparation can exert significant influence on the value of atmospheric electric field.

Apparently these data are not sufficient for direct exposure of this effect. Therefore we consider that for the efficiency of further researches, activity according to the preliminarily designed theoretical scenario might turn out useful. The present paper serves these objectives. Its main goal is strengthening of the physical base of a model of self-generated electromagnetic oscillations of the LAI system segment by the use of avalanche – like model of fault formation. Alongside with it we are offering here an attempt to check up the possibilities of atmospheric electric field, as a local indicator of the increasing of seismic activity.

## 2 Discussion

### 2.1 Avalanche-like unstable model of fault formation with the view of the theoretical model of self-generated seismo-electromagnetic oscillations of LAI system

The formula (1) (Kachakhidze, et all., 2011) connects with each other analytically the main frequency of the observed electromagnetic emission and the linear dimension (the length of the fault) of the emitted body:

$$\omega = \beta \frac{c}{l} \qquad (1)$$

where $\beta$ is the characteristic coefficient of geological medium and it approximately equals to 1. Of course it should be determined individually for each seismically active region, or for a local segment of lithosphere.

It should be emphasized that the avalanche-like unstable model (Fig.1) of fault formation that is well known in seismology (Mjachkin, et al., 1975) coincides with the model offered by us (Kachakhidze, et al., 2011) and explains well the given succession of EM emission changes in time taking place in the period of earthquake preparation: MHz, kHz and disappearance of emission directly before an earthquake (Johnston 1997., Eflaxias, et al, 2009).

It should be underlined that according to the model, on the basis of the electromagnetic emission in the period that precedes an earthquake it becomes possible to monitor the process of earthquake realization. Good conformity of the above considered two models is proved by the schemes of electromagnetic emission preceding the L'Aquila earthquake (Fig.2, 3) which are given in one of the papers (Papadopoulos, et al, 2010).

It is known that avalanche-like unstable model of fault formation is divided into three main stages (Fig.1): in case of large earthquakes the first stage can extend for dozen months; at this stage microcracks are generated chaotically, which have no definite orientation.

This stage of formation of microcracks is a reversible process - at this stage not only microcracks can be formed but also their so-called "locked one" can occur. Cracks created at this stage will be small (some dozen or hundred meter order) and their formation on the earth surface can be reflected in the form of weak or moderate foreshocks, which might not be close to the future earthquake epicenter. Thus, e.g. for the case of L'Aquila earthquake (fig.2) Dr. G. Papadopoulos states that "the present a posteriori analysis leaves no doubt that the state of seismicity before the L'Aquila main shock gradually changed from background seismicity to weak foreshock activity and then to strong foreshock sequence" (Papadopoulos, et al., 2010).

By the end of October 2008 the seismicity entered the state of weak foreshock sequence which lasted up to the 26 March 2009. It is characteristic that the weak foreshock activity which developed from 28 October 2008 to 26 March 2009 spatially did not concentrate around the mainshock epicenter but it was widely distributed within the seismogenic area (Papadopoulos, et al., 2010);

Such foreshocks can be conditionally referred to as the "regional foreshocks" (Kachakhidze, M. et al., 2003). Due to small length of microcracks and the reversibility of the process, the first stage in the electromagnetic emission frequency range, according to our model (Kachakhidze, et al., 2011) should be expressed in the discontinuous spectrum of MHz order emission (in radio diapason), which is proved (Eftaxias K., et al., 2009; Papadopoulos G., et al., 2010).

The second stage of the avalanche-like unstable model of fault formation is an irreversible avalanche process of the already somewhat oriented microstructures, which is accompanied by inclusion of the earlier "locked" ones. We have to suppose that this stage in the emission frequency spectrum should be expressed by MHz continuous spectrum already. According to the avalanche-like unstable model, this process takes place approximately several (at about 10-15) days before, which is evidenced by the material of observations over earthquakes (Papadopoulos, et al., 2010).

According to the avalanche-like unstable model, at the very stage gradual increase of cracks occurs (up to the kilometers order) at the expense of their uniting, to which, by formula (1), corresponds to the transition of MHz to kHz emission in the electromagnetic emission frequency spectrum.

Alongside with it, the relatively strong foreshocks can appear in seismogenic area too (Papadopoulos, et al., 2010; Wu et al., 1978; Wang, et al., 2006).

If a large earthquake is prepared and the $M \geq 5$ foreshock is not excluded it can well be that in the electromagnetic spectrum we'll have interchange of VLF and LF-frequencies, as it was in case of L'Aquila earthquake (Fig.2,3) (Papadopoulos, et al., 2010; Kachakhidze, et al, 2012).

At the final, third stage of the avalanche-like unstable model of fault formation the relatively big size faults use to unite into one, the main fault. This process, according to our model, in case of monitoring of emission spectrum should correspond to gradual fall of frequencies in kHz, which by formula (1) refers to the increase of fault length in the focus.

Increase of the fault length in the focus refers to the increase of the magnitude of the incoming earthquake. Fig.3 shows that by 4.04.2009 the main fault in the earthquake focus should have been of kilometer order.

Final formation of the main fault, that is relatively unstable zone, is accompanied by total fall of average "macro" stress in the greater part of the volume, since at this stage new faults are not created any more. It seems that this effect should be expressed in disappearance of EM emission that really takes place some hours before an earthquake. This fact is proved by observations on earthquakes (Johnston, 1997., Eftaxias, et al, 2009). The next, fourth stage in the avalanche-like unstable model of fault formation corresponds to the moment of an earthquake occurrence.

It should be emphasized that monitoring of EM emission frequency spectrum enables us to distinguish rather simply series of foreshocks and aftershocks from the main shock.

In the theoretical model (Kachakhidze, et al., 2011) which describes the possible mechanism of EM emission fixed in pre-earthquake period, EM emission was considered as earthquake indicator only, while on the basis of the above described analysis the precursory character of EM emission is vividly emphasized, since it turned out that it "brings" very precious information for prognostic diagnostic of the incoming earthquake.

## 2.2. Possible causes for anomalous changes of atmospheric electric field by the view of the model of self-generated electromagnetic oscillations of the LAI system segment.

According to the available theories, earth VLF electromagnetic emission generation might be conditioned by intensification of depth geophysical processes in the area of incoming earthquake preparation. [Eftaxias, et al, 2009; Eftaxias, et al, 2010]. Model of self-generated electromagnetic oscillations of LAI system (analogous contour), similar to other theories is based on the supposition that VLF electromagnetic emission is connected with depth processes going on in hypocenter area of incoming earthquake.

Usually, space space is considered as medium, where VLF electromagnetic emission is revealed only in the form of disturbing factor. For example, according to the LAI segment self generated electromagnetic oscillations model, the initial exposure of depth phenomena is polarization of the  layer of earth surface, which itself will not generate VLF electromagnetic emission. For it, polarization should be accompanied by induction effect, in order to contribute somehow to the change of polarity signs between the earth surface and atmosphere (lower ionosphere). Later this zone will become a zone of generation of VLF electromagnetic waves. It is quite natural to propose that self-generated electromagnetic oscillations will cause local changes in the energetic picture corresponding to the given segment of LAI system. As a result of it, the parameters which together with frequency-amplitude characteristics form conditions for spreading and attenuating of electromagnetic disturbance, will suffer changes. One of the determining factors of these conditions should be the Earth- lower ionosphere  global waveguide, which functions efficiently in night hours and contributes to channelized spreading of electromagnetic waves at rather long distances. It is known that spreading of (3-30 kHz ) frequency range VLF electromagnetic waves accompanying atmospheric electric discharge is connected with the action of this waveguide. Therefore it is logical that in some cases waveguide of  earth-ionosphere should make effect on spreading of earth VLF electromagnetic emission too.

Within the frames of formalism of the analogous contour a cause that conditions change of polarization sign of the local section of the earth surface has no principal significance. This can be a piezo-effect as well as the depth electro-kinetic process conditioned by increase of seismic activity.

The change of polarity sign between the earth surface and lower ionosphere implies that in analogous contour zone we'll have inversion of a sign of a gradient of the potential of natural atmospheric electric field. Therefore in the period that precedes earthquake intensification of inversions phenomenon, together with VLF electromagnetic emission might be considered as a sign of development of electromagnetic induction effect generated after polarization.  Although in the ionosphere medium generation of electromagnetic waves and their attenuating in time can proceed so non-homogeneously and by irregular discretion, that atmospheric electric field inversion will not be fixed. Alongside with it, we must take into consideration that association of inversion effect definitely with seismic activity (that is with earthly source) without exclusion of space disturbance factors of atmospheric electromagnetic field would not be correct. Therefore, for determination of concrete reason of inversion and for adequate interpretation of the data of observations qualitative energetic evaluation of the process of spreading of wave disturbance in ionosphere medium might be useful, at the support of analogous contour formalism.

Namely, a zone of functioning of the earth-lower ionosphere waveguide contributing to spreading of disturbance in night hours practically coincides with the probable zone of analogous contour. One more fact should be stated here: the main (critical) frequency of  ideal ionosphere waveguide, similar to frequency of analogous contour, is determined by its linear scale, waveguide height, that is distance from the earth to the lower ionosphere (Lichtner et al 1988)

$$\omega_{cr.} = \frac{c}{2h} \qquad (2)$$

where h-is height, $\omega_{cr.}$ – critical frequency.

Thus, if the value of electromagnetic waves frequency is lower than the definite value, their canalized spreading in waveguide channel will be impossible. And since there are definite quantitative limits it might be assumed that near the critical frequency in waveguide we will have generation of standing electromagnetic waves. It is not excluded that generation of standing waves will take place in analogous contour area too, similar to waveguide. In such case, we can use analogy with common linear conductors, between which, at definite conditions it will be possible to generate standing electromagnetic waves [Shimoni K.,1964].

In distinct from the common electromagnetic waves, energy of standing waves is localized in the area of their generation. Therefore, their existence will cause oscillation of tension vector of atmospheric electric field with the self-generated frequency of analogous contour. Although in distinct from the contours created from linear conductors, complete blocking of analogous contour, which is a prerequisite of generation of standing electromagnetic waves, is hardly imaginable. Respectively, there is only theoretical possibility of resonance strengthening of electromagnetic waves in analogous contour zone. For example, it may happen only in the hypocentral area of incoming earthquake, especially in the process of formation of main fault.

According to the formula (2) only in this case the self –generated electromagnetic wave frequency of analogous contour will approximate to Earth-lower ionosphere waveguide critical frequency. At this moment extremely sensible fall of electromagnetic energy stream is expected. Because of it, in the space near the analogous contour a decrease of spectral weight corresponding to waveguide critical frequency should be expected in the whole electromagnetic emission. At the support of this effect which can develop in analogous contour area, decrease of intensity of VLF electromagnetic emission that corresponds to earth-lower ionosphere waveguide critical 1.7 kHz frequency in night hours preceding the earthquake might be explained rather simply.

According to this hypothesis for explanation of the above stated phenomenon alteration of waveguide height might be not necessary [Nˇemec, et al., 2009]. It is natural that VLF electromagnetic waves frequency of which exceeds the critical value can leave the area of analogous contour and spread towards waveguide. In this way the transfer of electromagnetic energy which incites disturbance of weakly ionized ionospheric medium is realized.

Theoretically, in case of increase of size of analogous contour, that implies decrease of frequency of VLF electromagnetic wave frequency, or correspondingly, decrease of energy density, intensity of disturbance should fall. Although in case of open analogous contour it can't be excluded that VLF electromagnetic waves with relatively lower frequency will cover rather wide area of ionosphere [Kachakhidze, et al., 2011]. In this case density of electromagnetic energy stream at some distance from analogous contour may correspond in size to the characteristic energy density of epicenter area of incoming earthquake. In this case, anomalous changes of the value of atmospheric electric field might be fixed in some distance from the incoming earthquake epicenter area too. Therefore to check up this opinion we consider appropriate to study the picture of peculiarities of anomalous changes of atmospheric electric field value by means of concrete examples.

It is natural that for correct resolution of this task all irregular sources which cause disturbance of atmospheric electric field, without a factor of increasing of seismic activity, should be excluded. In addition, in this process, energetic effects of running and standing VLF electromagnetic waves could be separated from each other.

Thus, proceeding from the properties of standing electromagnetic waves, in case of their generation in analogous contour medium, the oscillation of gradient of atmospheric electric field should be of harmonious character. Therefore, irrespective of the fact that in standing wave nodes instant value of amplitudes of electric field tension might be rather big, in the process of field averaging this effect might be insignificant.

In other case, that is when we'll have generation of running VLF electromagnetic waves, it is probable that earth-ionosphere waveguide contributes to spreading of these waves at a rather long distances. At this moment especially active will be energy transfer in ionosphere medium and development of dissipation effect, which should be immediately expressed on the value of atmospheric electromagnetic field and its polarity sign.

To clarify, as much as possible, the above referred problems, let's consider the information of multiyear records about atmospheric electric field obtained by us from Geophysical Observatory of Dusheti (Georgia). This place is known by especially stable cosmo-geophysical characteristics within the range of the whole South Caucasus region ( $\varphi = 42^0 05'; \lambda = 44^0 42'$ ).

Alongside with it, Dusheti is characterized by high number of sunny days ($\approx 74$ %) and respectively, few days of bad weather. Because of it, rather large mass of the multiyear data of this observatory (1953-92) was used by us for checking up  the efficiency of atmospheric electric effect average and large earthquakes ($M \geq 4.5$) which happened in Caucasus (within 40- 600 km from Dusheti),  as one of the earthquake indicators  [Kachakhidze, et al., 2009].

As a result of analysis performed in this work, a conclusion was made that within 11 day interval that precedes earthquake there is a rather great probability of anomalous changes in atmospheric electric field.

Thus, it is interesting to made clear if the intensity of disturbance of atmospheric electric field depends on incoming earthquake magnitude and on the distance from the point of observation. Available studies fail to provide us with more or less reliable answer to this question yet. Therefore we consider useful to address again the early used base of data. This time the analysis aims to clear up the character of behavior of atmospheric electric field at the stationary point of observation.

The below given figures are illustrations of tendencies revealed in anomalous changes of gradient of atmospheric electric field. Figure 4, as the first example, offers dynamic synthesized picture of changes in atmospheric electric field, which was construed by the data of 1970. This picture is a result of superimposition of time intervals data corresponding to some seismically less active periods (at about 10 day periods). Selection of these intervals was performed according to the following criteria: in the selected periods earthquake of $M \geq 3.6$ were not fixed, although lower magnitude earthquakes happened rather often.

On the Fig.4 and on all other figures time on the axis of abscissas is measured in hours, while on the axis of ordinates -10-foldreduced gradient of atmospheric electric field potential   is given in V/m. Likewise, on all figures zero level of electric field correspond to reference significance adopted by Dusheti Observatory: 84 V/m, which was determined by averaging of multiyear data. It is apparent that  all sources of disturbance of atmospheric electric field are contributing to the background synthesized picture of  Fig.4. Although due to low seismic activity within time intervals corresponding to background significances, we can assume that effect of earth VLF electromagnetic emission on this picture is insignificant. Apparently change in polarity of a background significance of atmospheric electric field occurs rather rarely, while limits characteristic to changes of this index are rather narrow (mainly some hours and very seldom one or two days).

In the period of seismic activity that is increased compared to the background level, which ends by average or large  ($M \geq 4.5$)   earthquakes, probably the intense earth VLF electromagnetic emission generation took place. It is quite natural that this effect could contribute to a picture of irregular disturbance of atmospheric electric field. Among the sources of atmospheric electric field disturbance alongside with the VLF electromagnetic waves, there are meteorological factors and space (magnetosphere) VLF electromagnetic emission. The latter seems rather discrete and its intensity is in direct relation with the level of geomagnetic field disturbance [Kachakhidze, et al., 2012].

It is known that this parameter determines greatly irregular changes of meteorological situations on the earth. Therefore, if we exclude irregular meteorological factor in a picture that expresses atmospheric electric field, it will be rather correct to consider that similarly, the magnetospheric VLF emission effect will be excluded too. In this case, we may consider that the remainder of irregular disturbance of atmospheric electric field is connected only with the earth VLF electromagnetic emission, the generation area of which is localized in epicenter zone of an incoming earthquake.

Thus, our task deals with local problems of the scales of the Caucasus region. Atmospheric electric field one-hour records are considered for minimum 10 days before earthquake, for which $\bar{x} \pm 3\sigma$ approximation was used. Here $\bar{x}$ is seasonal average significance of atmospheric electric field, while $\sigma$ is standard deviation [Kachakhidze, et al., 2009]. Application of such criterion implies that from the changes in atmospheric electric field any natural and anthropogenic irregular disturbing factor is practically excluded, which was an objective of our task. For a concrete analysis we have selected some wide groups of the same magnitude earthquakes which happened at various distances from Dusheti (from M=3.6 to M=6.9) Analysis of the data of these earthquakes processed by the above stated accuracy statistical method, and the obtained results revealed that the level of atmospheric electric field disturbance compared to that of the background, as a rule increases. Namely the probability of inversion of electric field polarity increases.

This effect turned out of clearly irregular character and similar to amplitudes characteristic of electric field intensity, depend neither on earthquake magnitude nor on a distance form place of observation. For illustration of these results Fig. 5,6,7 offer pictures of disturbance of atmospheric electric field potential gradients in the periods corresponding to two earthquakes of average (M=4.0) and large (M=6.9) magnitudes.

Similar result is proved by synthetic pictures of atmospheric electric field potential gradients corresponding to the same magnitude earthquakes. Namely, Fig 7 offers curve corresponding to M=4.3 earthquakes; its comparison with background curve (Fig.4) clearly refers to the fact that the atmospheric electric field disturbance level increases in the period that precedes earthquake preparation. But similar to this figure, in other figures, isolation of regular element in the behavior of atmospheric electric field is void of grounds. It is quite natural that such result incites doubt in efficiency of atmospheric electric field gradient, as a reliable electric indicator of earthquake preparation process.

It should be stated that practically the same opinion is fixed by Pulinets [Pulinets, et al., 2004], although the mass of our data is greater and what is most important, it belongs to only one region. It means that electric field anomalies are spread practically in the same parameter ionosphere medium.

**Conclusions**

1. On the base of the model of self-generated seismo-electomagnetic oscillations of the LAI system the avalanche-like unstable model of fault formation explains well succession of changes in frequency of the existing electromagnetic spectrum of VLF emission, in the period that precedes the earthquake;

2. As a result of processing of atmospheric electric field potential gradient data it turned out that the increase of seismic activity in the Caucasus region is characterized by increase of atmospheric electric field inversion, although this effect is not in correlative relation with incoming earthquake magnitude and location of epicenter; therefore, according to our data atmospheric electric field gradient can not be considered as a reliable electric indicator for the process of earthquake preparation;

3. By the use of analogous contour model, decrease of intensity of VLF electromagnetic emission with critical 1.7 kHz frequency for earth – lower ionosphere waveguide during night hours before earthquakes might be explained easily.

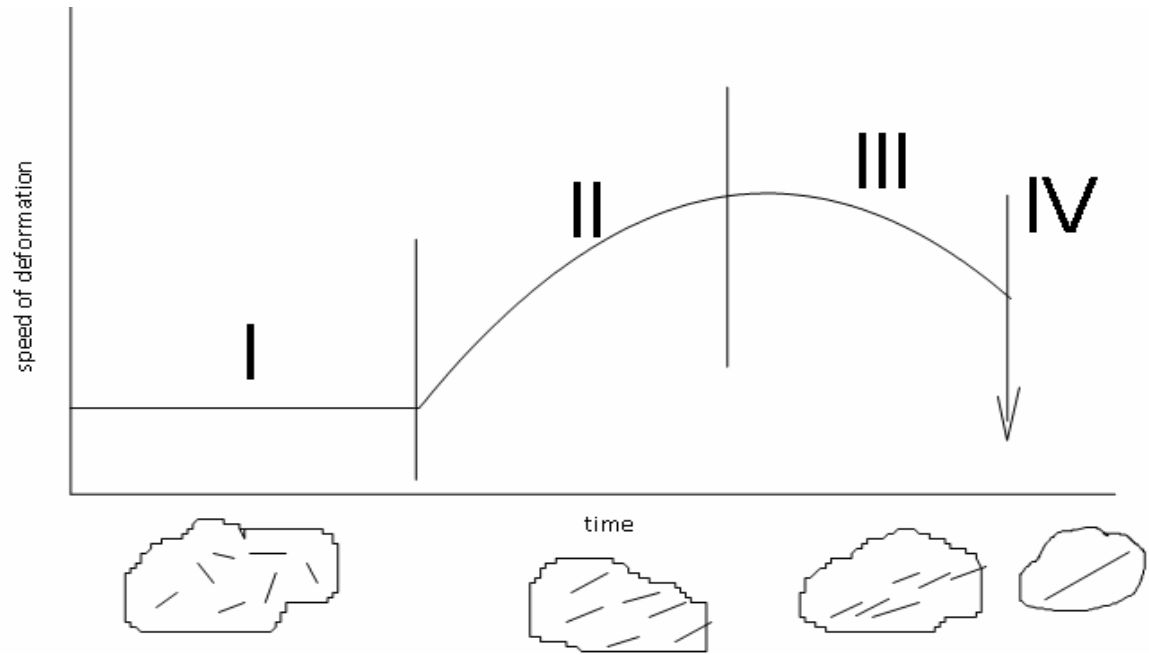

Fig. 1. Scheme of fall-unstable model of fracture origination

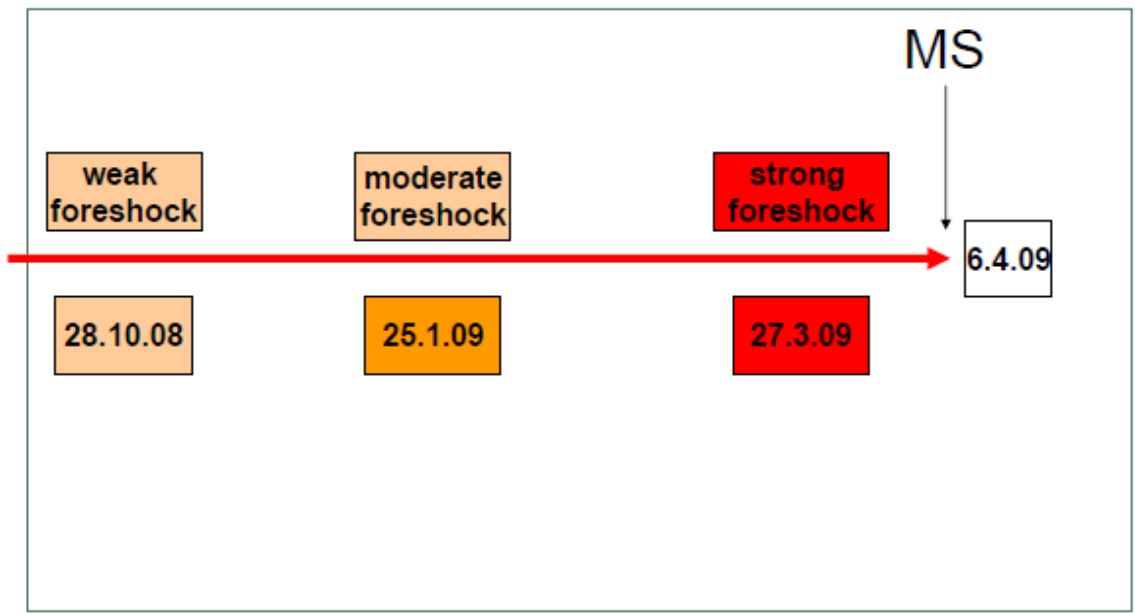

Fig. 2. Evolution of foreshock activity

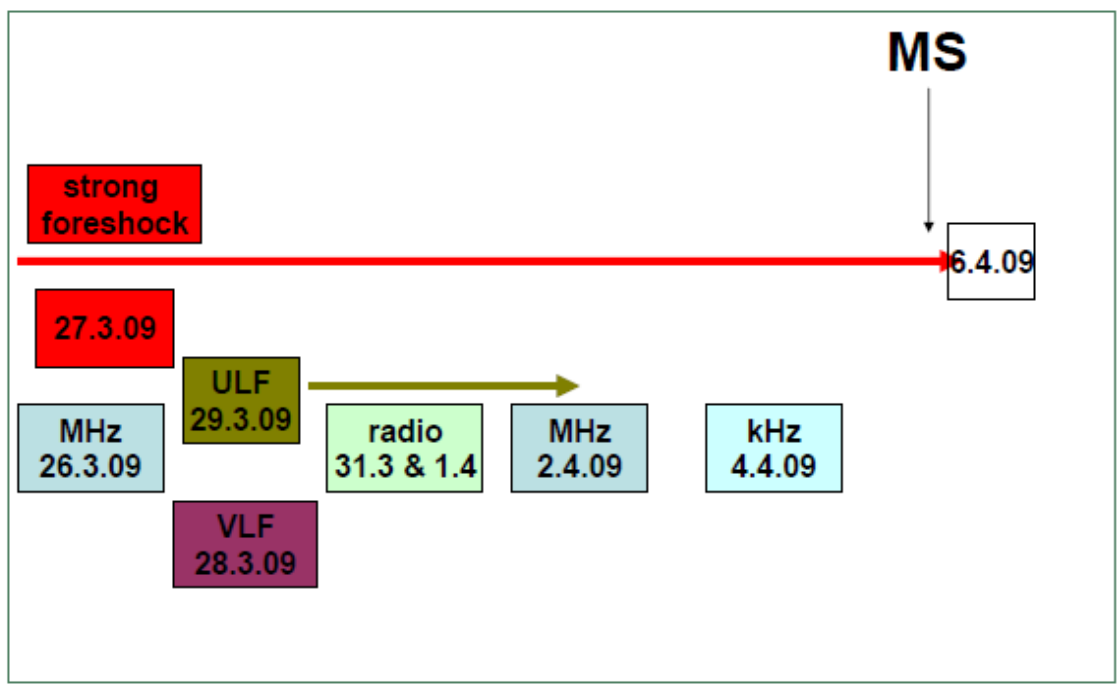

Fig. 3 Evolution of EM emission

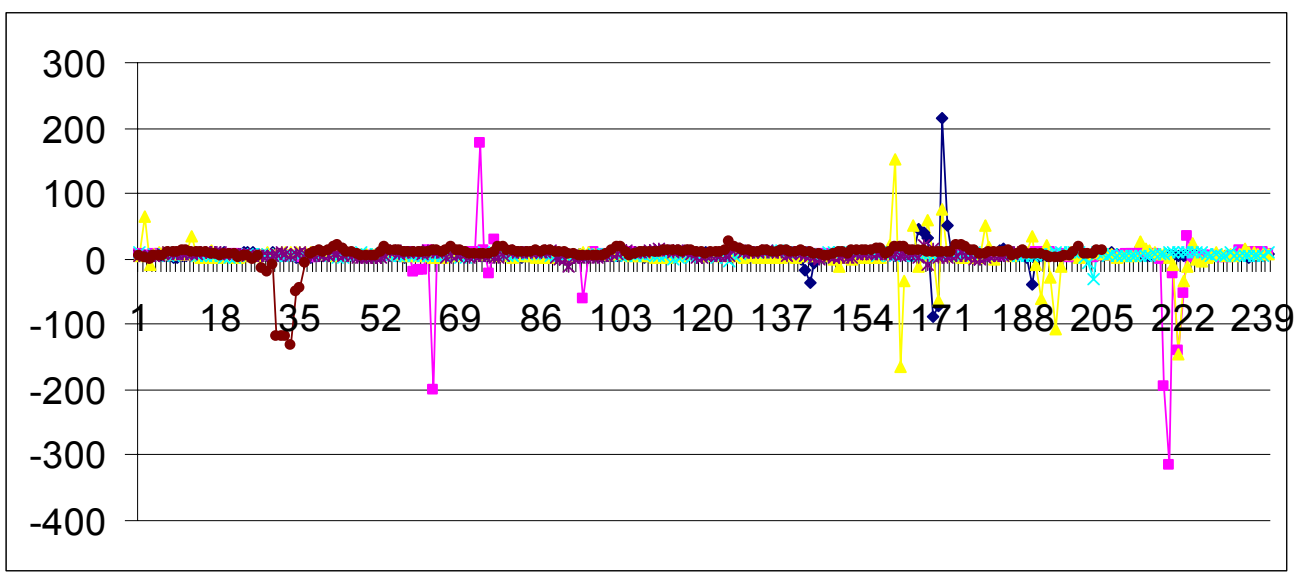

Fig. 4 . Synthesized picture of changes of atmospheric electric field potential gradient in seismically relatively quiet periods of 1970

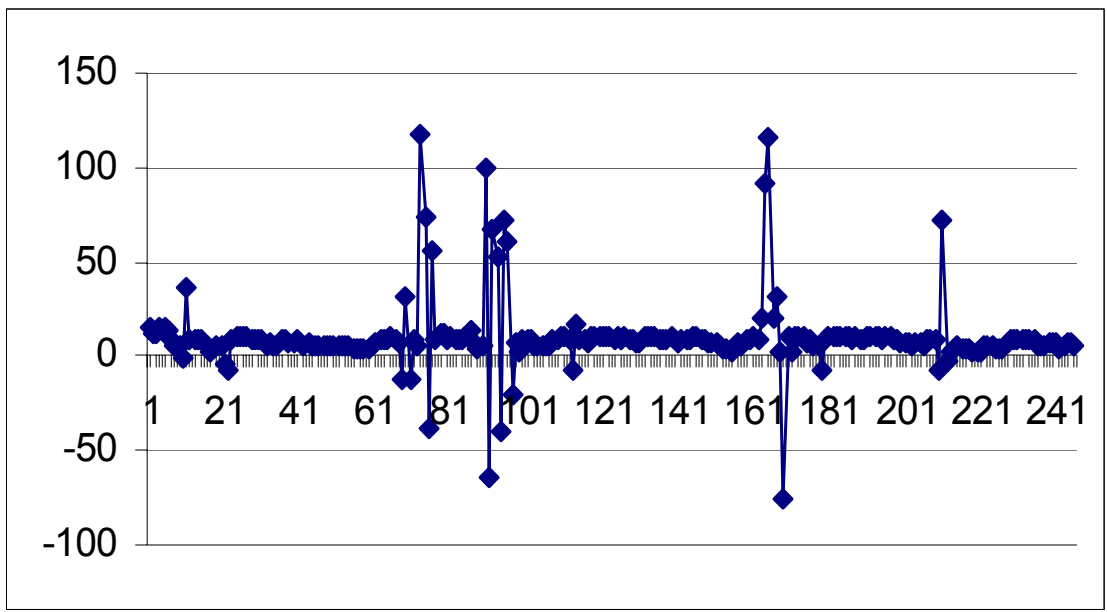

Fig. .5 . Changes in atmospheric electric field potential gradient before  M=4.0  earthquake (1970. 04. 21; distance from Dusheti 441 km)

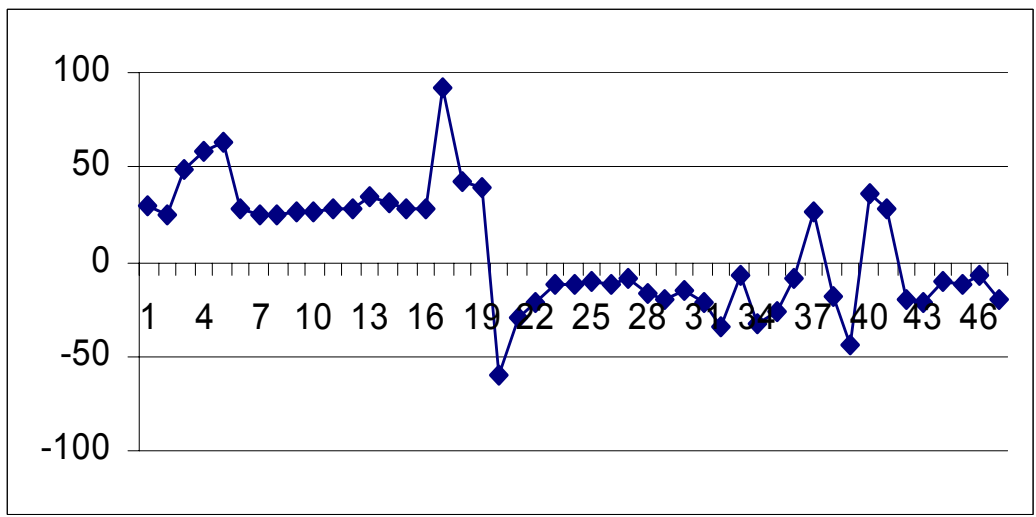

Fig. 6. Atmospheric electric field potential gradient changes in the period preceding   M=6.9 earthquake (1991. 04. 29; Distance from Dusheti 72 km.)

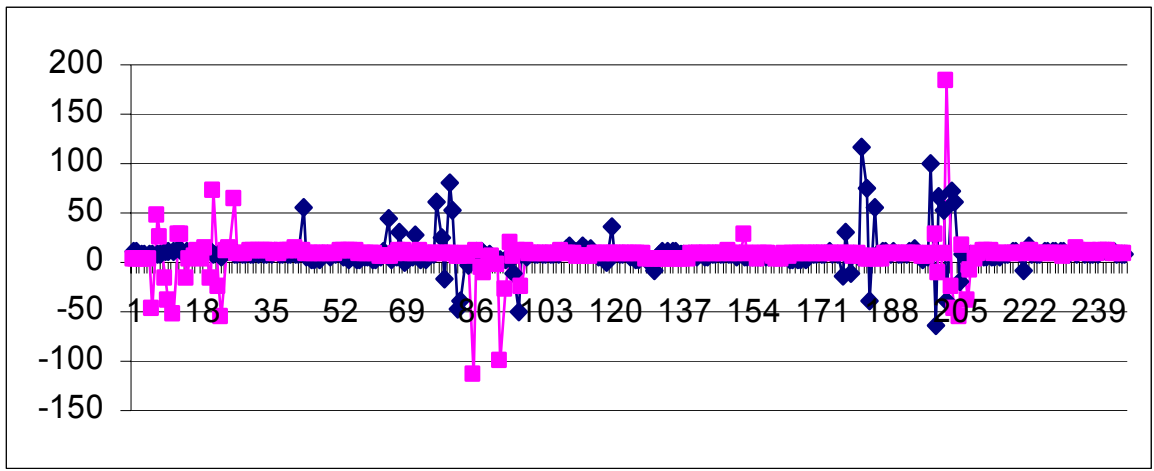

Fig. 7. Synthetic picture of changes of atmospheric electric field potential gradient in the period that preceded M=4. 3 earthquake